\documentclass[aps,prl,twocolumn,epsfig,floats]{revtex4}
\usepackage{graphicx,epsf,natbib,amsmath,latexsym,amssymb}
\def\be{\begin{equation}}
\def\ee{\end{equation}}
\def\bea{\begin{eqnarray}}
\def\eea{\end{eqnarray}}

\begin{document}


\author{Huazhou Wei, Sung-Po Chao, Vivek Aji}
\affiliation{Department of Physics 
and Astronomy, University of California, Riverside, CA 92521}
\title{Excitonic Phases from Weyl Semi-Metals}

\date{\today}

\begin{abstract}
Systems with strong spin-orbit coupling, which competes with other interactions and energy scales, offer a fertile playground to explore new correlated phases of matter. Weyl semimetals are an example where the phenomenon leads to a low energy effective theory in terms  of massless linearly dispersing fermions in three dimensions. In the absence of interactions chirality is a conserved quantum number, protecting the semi-metallic physics against perturbations that are translationally invariant. In this letter we show that the interplay between interaction and topology yields a novel chiral excitonic insulator. The state is characterized by a complex vectorial order parameter leading to a gapping out of the Weyl nodes. A striking feature is that it is ferromagnetic, with the phase of the order parameter determining the direction of the induced magnetic moment.
\end{abstract}
\maketitle

The emergence of materials whose properties are strongly influenced by spin orbit coupling is an exciting new phenomenon in condensed matter physics. Topological insulators \cite{Kane,Roy} and Weyl semi-metals\cite{Savrasov,Balents1,Burkov} are two canonical examples, where the degrees of freedom, describing the low energy physics of a non-relativistic many body system, are linearly dispersing massless fermions. An interesting possibility in these materials is the emergence of new phases of matter\cite{Balents2,Ran}. In this letter we focus on the Charge Density Wave (CDW) and Excitonic Insulating (EI) states in Weyl semi-metals. 

Weyl fermions are conjectured to be the low energy excitations of Pyrochlore Irradiates (PIs)\cite{Savrasov} and Topological-Normal insulator (TNI) heterostructures\cite{Balents1,Burkov}. The semi-metallic nature is derived from the touching of two non degenerate bands at an even number of Weyl points. The existence of such points leads to a number of anomalies in transport characteristics\cite{Ninomiya,Aji,Vishwanath}. While the physics of Weyl fermions has been extensively studied in the context of liquid $^{3}$He\cite{Volovik,Tsuneto}, where they arise in the $A$ phase, the recent developments have renewed searches for other systems that realize the phenomena\cite{Berlinsky}.

An important property of Weyl points is that they are robust to most interactions. This is a consequence of chirality being a conserved quantum number in three dimensions for massless fermions. Only interactions that couple fermions of opposite chirality can open a gap. For systems that preserve inversions but break time reversal, every Weyl node, if it exists, is accompanied by another of opposite chirality in the Brillouin zone. In the case of PIs 24 nodes at symmetry related points are conjectured to exist\cite{Savrasov} while the TNI heterostructures have two such nodes\cite{Balents1}. In both cases there is perfect nesting. Consequently we seek to address the question of the possible particle-hole instabilities promoted to by repulsive interactions. 

To highlight the physics of the novel new state, we simplify to the case of two Weyl nodes and local density density interactions. A more general analysis including the effects of long range Coulomb will be reported elsewhere. There are two types of particle-hole excitations that can arise in this case i) intra-nodal (occurring at zero momentum) and ii) inter-nodal (occurring at a finite fixed momentum associated with the nesting vector). These are excitonic phases\cite{Wk}, the former being the EI while the latter is the CDW. Unlike conventional condensates in these sectors, the electron hole pairing of Weyl fermions leads to chiral phases. A minimum interaction strength is required to nucleate them which is the consequence of the vanishing density of states at the node. For local interactions, the chiral EI has the lowest threshold, opens a gap at the nodes and is the most stable state. The sign of the gap is opposite at the two nodes, preserving inversion symmetry. 

The order parameter breaks $SU(2)\times U(1)$ symmetry corresponding to rotational and phase invariance. This is in addition to the time reversal symmetry broken by the parent state. To characterize the broken symmetry state we define an orthonormal basis for three dimensional space $\{\hat{l}, \hat{m}, \hat{n}\}$. In particular, for the cartesian coordinate system with $\hat{z}$ as the quantization axis, the ordered state is one with $\left<\sum_{\vec{k}}\hat{e}_{\vec{k}}^{2}\left(c_{\vec{k},+}^{L\dag}c_{k,-}^{L}-c_{\vec{k},+}^{R\dag}c_{k,-}^{R}\right)\right>$ $=\Delta \left(\frac{\hat{x}\pm \imath \hat{y}}{\sqrt{2}}\right) \exp\left(\imath\chi\right)$, where $c_{\vec{k},\pm}^{L,R}$ is the fermion annihilation operator at the two Weyl nodes, labeled $L$ and $R$, at momentum $\vec{k}$ in the band labeled $\pm$, $\hat{e}_{\vec{k}}^{2}= \{-\hat{k}_{y}/\sqrt{\hat{k}_{x}^{2}+\hat{k}_{y}^{2}}, \hat{k}_{x}/\sqrt{\hat{k}_{x}^{2}+\hat{k}_{y}^{2}},0\}$, $\Delta$ is the magnitude and $\chi$ is the phase of the order parameter. The loss of rotational invariance is reflected in the the vector $\hat{e}_{\vec{k}}^{2}$ appearing in the order parameter. In general it is a vector lying in the plane spanned by $\hat{l}$ and $\hat{m}$ (i.e. $\hat{e}_{\vec{k}}^{2}= \hat{n}\times \hat{k}/\left|\hat{n}\times \hat{k}\right|$ with $\hat{n} = \hat{l}\times\hat{m}$) and the corresponding order parameter is $\Delta \left(\frac{\hat{l}\pm \imath \hat{m}}{\sqrt{2}}\right) \exp\left(\imath\chi\right)$. 

The order parameter does not break inversion symmetry. Therefore the spin orientation for electrons at momentum $\vec{k}$ and $-\vec{k}$ are identical. The effect of the symmetry breaking is to cant spins in momentum space along a direction determined by the phase of the order parameter. Thus vortex lines, if stabilized, in this system have an associated spin texture with finite divergence, and the singularity localized in the vortex core.

\noindent\underline{\textit {Model}}: Consider a system with two Weyl nodes at $\vec{K}_{1} = K_{0}\hat{x}$ (labeled R) and $-\vec{K}_{1}= -K_{0}\hat{x}$ (labeled L) with chiralities $+1$ and $-1$ respectively. The Hamiltonian is

\begin{equation}
H_{0 \pm} = \pm \hbar v \sum_{\vec{k}}\psi_{\vec{k}\alpha}^{\dagger}\vec{\sigma}_{\alpha\beta}\cdot\left(\vec{k}\mp \vec{K}_{0}\right)\psi_{\vec{k}\beta}
\end{equation}
where $v$ is the fermi velocity and $\vec{\sigma} = \{\sigma_{x}, \sigma_{y}, \sigma_{z}\}$ is a vector of Pauli matrices. The dispersion at each node is $\epsilon_{\vec{q}}=\pm \hbar v\left|\vec{q}\right|$ centered around $\pm K_{0}$, with $\vec{q} =\left(\vec{k}\mp \vec{K}_{0}\right)$. The conduction (valence) band at the R node has its spin parallel (anti-parallel) to $\vec{q}$, while the opposite is true at the L node.
The general particle particle interaction takes the form
\begin{eqnarray}\nonumber
V&=&\sum_{\sigma,\sigma'}\int d\vec{r} d\vec{r}' V\left(\vec{r}-\vec{r}'\right)\psi^{\dagger}_{\sigma'}\left(\vec{r}'\right)\psi_{\sigma'}\left(\vec{r}'\right)\psi^{\dagger}_{\sigma}\left(\vec{r}\right)\psi_{\sigma}\left(\vec{r}\right) \\\label{vg}
&=&\sum_{\sigma,\sigma'}\sum_{\vec{k}, \vec{k}',\vec{q}}V(\vec{q}) \psi^{\dagger}_{\vec{k'}+\vec{q},\sigma'} \psi_{\vec{k'},\sigma'}\psi^{\dagger}_{\vec{k}-\vec{q},\sigma}\psi_{\vec{k},\sigma}
\end{eqnarray}
For the moment we do not make any assumptions on the nature of the interactions. Since the Weyl physics is the low energy description of a more general theory, we enforce an upper cutoff in the momentum integrals (up to an energy $\Lambda$) around the Weyl point. The precise source of the interaction and the renormalization effects will be considered elsewhere. Here we focus on the instabilities within a mean field analysis.\\

\begin{figure}
\includegraphics[width=0.5\columnwidth, clip]{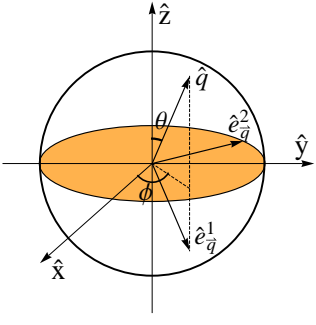}
\caption{The interaction shown in Eq.(\ref{int}) is a function of three vectors ($\hat{q}$, $\hat{e}^{1}_{\vec{q}}$ and $\hat{e}^{2}_{\vec{q}}$) that form a right handed coordinate system. Each vector couples to an operator of distinct symmetry in the particle hole channel.}
 \label{vectors}
\end{figure}

\noindent\underline{\textit {Particle Hole instabilities}} :  We rewrite the interaction in the basis of the non-interacting bands. Define $\psi_{\vec{q},\pm}^{R,L} = \eta_{\vec{q},\pm}^{R,L}c_{\vec{q},\pm}^{R,L}$ as the fermonic field of the four bands, where $\eta$ is the spinor and $c$ is the fermion annihilation operator.  Of the $16$ possible terms from Eq.(\ref{vg}) only $6$ terms satisfy momentum conservation for scattering restricted to the states within the cutoff around the node. 

For every momentum $\vec{q}=q\hat{q}$, where $\hat{q} = \{\hat{q}_{x}, \hat{q}_{y}, \hat{q}_{z}\}$ is the unit vector along $\vec{q}$, we define two orthogonal vectors $\hat{e}_{\vec{q}}^{1} \equiv \hat{\theta}_{\vec{q}} = \{\hat{q}_{x}\hat{q}_{z}/\sqrt{\hat{q}_{x}^{2}+\hat{q}_{y}^{2}},\hat{q}_{y}\hat{q}_{z}/\sqrt{\hat{q}_{x}^{2}+\hat{q}_{y}^{2}}, -\sqrt{\hat{q}_{x}^{2}+\hat{q}_{y}^{2}}\}$ and $\hat{e}_{\vec{q}}^{2}\equiv \hat{\phi}_{\vec{q}} = \{-\hat{q}_{y}/\sqrt{\hat{q}_{x}^{2}+\hat{q}_{y}^{2}}, \hat{q}_{x}/\sqrt{\hat{q}_{x}^{2}+\hat{q}_{y}^{2}},0\}$, such that $\hat{q}$, $\hat{e}_{\vec{q}}^{1} $ and $\hat{e}_{\vec{q}}^{2} $ form a right handed coordinate system (see Fig.\ref{vectors}). The unit sphere is spanned by the vector $\hat{q}$ by two rotations, one about any axis perpendicular to $\hat{e}^{2}_{\vec{q}}$ and the another about $\vec{e}_{\vec{q}}^{2}$. For example if we choose the first to be the $z$-axis, than the vector $\vec{e}^{2}_{\vec{q}}$, which is the is $\hat{\phi}$  in the spherical polar system, spans a unit circle (perimeter of the shaded region in Fig.\ref{vectors}) and the vector $\vec{e}^{1}_{\vec{q}}$, which is the corresponding $\hat{\theta}$, spans the southern hemisphere.  The following construction holds for an arbitrary quantization axis $\hat{n}$, with the corresponding polar and azimuthal angle for $\vec{q}$ defined in the coordinate frame  $\{\hat{l}, \hat{m}, \hat{n}\}$. In the rest of the letter we use the $\{\hat{x}, \hat{y}, \hat{z}\}$ coordinate system.

Specializing to potentials that are even functions of $\vec{k}$, i.e. $V(\vec{k}) = V(-\vec{k})$, the interaction in terms of $\hat{e}_{\vec{k}} = \hat{e}_{\vec{k}}^{1}+\imath \hat{e}^{2}_{\vec{k}}$, is  
\begin{widetext}
\begin{eqnarray}\nonumber
V=&-&\sum_{\vec{k},\vec{k}',n=\pm}\left[ {V(\vec{k}-\vec{k}')\over 4}\left(\hat{e}_{\vec{k}}\cdot \hat{e}_{\vec{k}'}^{\ast}+ \hat{e}_{\vec{k}} ^{\ast}\cdot\hat{e}_{\vec{k'}} \right)\sum_{\tau=R,L}c_{\vec{k},n}^{\tau\dag }c_{\vec{k},-n}^{\tau}c_{\vec{k}',-n}^{\tau\dag}c_{\vec{k}',n}^{\tau}+{V(\vec{k}-\vec{k'}-2\vec{K}_{0})\over 2}\left(\hat{e}_{\vec{k}} \cdot\hat{e}_{\vec{k}'}+ \hat{e}_{\vec{k}}^{\ast}\cdot \hat{e}_{\vec{k}}^{\ast} \right)\right.\\\label{int}
&\times&\left. c_{n\vec{k}}^{L\dag}c_{-n\vec{k}}^{L}c_{-n\vec{k}'}^{R\dag}c_{n\vec{k}'}^{R}-\left(2V(2\vec{K}_{0})-{V(\vec{k}-\vec{k}')}\left(\hat{k}\cdot\hat{k}'+1\right)\right) c_{n\vec{k}}^{L\dag}c_{-n\vec{k}}^{R}c_{-n\vec{k}'}^{R\dag}c_{n\vec{k}'}^{L}\right]
\end{eqnarray}
\end{widetext}
The first and second term promote CDW instabilities with intra-nodal order parameter, i.e. $\left<\sum_{\vec{k}}\vec{A}_{\vec{k}}c^{\tau\dag}_{n\vec{k}}c^{\tau}_{-n\vec{k}}\right>\neq 0$ with $\vec{A}_{\vec{k}}$ an odd function of $\vec{k}$, while the last term leads to inter-nodal order with $\left<\sum_{\vec{k}}\vec{A}_{\vec{k}}c^{\tau\dag}_{n\vec{k}}c^{\bar{\tau}}_{-n\vec{k}}\right>\neq 0$.

\noindent\underline{\textit {Inter-nodal Charge density wave}} : We begin by studying the inter nodal instability that establishes ordering at $2\vec{K}_{0}$ (third term in Eq.(\ref{int})). For momentum independent interaction potentials, $V(\vec{k}) = g/\Omega$, where $\Omega$ is the volume of the system, the coupling takes the form
\begin{equation}\label{veff}
V_{eff}=-{g\over \Omega}\sum_{\vec{k},\vec{k}'} \sum_{n=\pm}\left(\hat{k}c_{n\vec{k}}^{L\dag}c_{-n\vec{k}}^{R}\right)\cdot\left(\hat{k}'c_{-n\vec{k}'}^{R\dag}c_{n\vec{k}'}^{L}\right)
\end{equation}
Eq.(\ref{veff}) is identical to that of the interaction in $^{3}$He in the particle particle channel that leads to chiral superfluidity\cite{Volovik,Tsuneto}. In Weyl semi-metals the corresponding state in the particle-hole channel is a charge density wave. Within mean field there are two possible instabilities: i)  Chiral CDW: $\vec{\Delta}_{c}=\frac{g}{\Omega}\left<\sum_{\vec{k'}}\hat{k'}c^{\tau\dag}_{n\vec{k'}}c^{\bar{\tau}}_{-n\vec{k'}}\right>=\Delta_{c}( \frac{\hat{x}+\imath \hat{y}}{\sqrt{2}})$ and ii) Polar CDW:  $\vec{\Delta}_{p}=\frac{g}{\Omega}\left<\sum_{\vec{k'}}\hat{k'}c^{\tau\dag}_{n\vec{k'}}c^{\bar{\tau}}_{-n\vec{k'}}\right>=\Delta_{p} \hat{z}$. Note that the directions chosen for the ground state is for convenience and all other that are obtained by rotation in three dimensions are equivalent. The former is a chiral state while the latter is a non-chiral p wave CDW. The general form of mean field Hamiltonian is
\begin{eqnarray*}
H_{MF} = \sum_{\vec{q},n}\left(\begin{array}{c} c_{\vec{q},n}^{L} \\ c_{\vec{q},-n}^{R} \end{array}\right)^{\dagger} \left(\begin{array}{cc} \hbar v n|\vec{q}| & -\vec{\Delta}\cdot\hat{q}\\ -\vec{\Delta}^{\ast}\cdot\hat{q} & -\hbar v n|\vec{q}| \end{array}\right) \left(\begin{array}{c} c_{\vec{q},n}^{L} \\ c_{\vec{q},-n}^{R} \end{array}\right)
\end{eqnarray*}
with $\vec{\Delta}$ representing $\vec{\Delta}_c$ or $\vec{\Delta}_p$.
Minimizing the free energy with respect to the order parameter, the integral equation that determines the condition on the coupling constant leading to the instability is
\begin{equation}\label{gap}
1 = {g\over {\Omega}}\sum_{\vec{k}}{|\hat{\Delta}_{c/p}\cdot\hat{k}|^{2}\over{2E_{\vec{k}}}}\tanh(\beta E_{\vec{k}}/2)
\end{equation}
Here $E_{\vec{k}}=\sqrt{(\hbar v|\vec{k}|)^2+|\vec{\Delta}\cdot\hat{k}|^2}$, $\beta=1/k_B T$, $\hat{\Delta}_{c/p}$ represents the unit vector direction of chiral or polar state, and $\Omega=\frac{8\pi}{3n}\left(\frac{\Lambda}{\hbar v}\right)^3$, where $n$ is the number of electrons in the system, set to $1$ in the following. The summation over momentum is cutoff by the scale $\Lambda/\hbar v$. In the limit of $\Delta \ll \Lambda$ and zero temperature, to leading order Eq.(\ref{gap}) simplifies to
\begin{eqnarray*}
&1&\simeq \frac{\pi g \Lambda^2}{(\hbar v)^3}\left(\frac{2}{3}+\left(\frac{-17-30\ln2}{225}+\frac{4\ln\left(\frac{\Delta_{c}}{2\Lambda}\right)}{15}\right)\frac{\Delta_c^2}{\Lambda^2}+\ldots\right)\\
&1& \simeq\frac{\pi g \Lambda^2}{(\hbar v)^3}\left(\frac{2}{3}+\left(\frac{3}{25}+\frac{2}{5}\ln\left(\frac{\Delta_p}{2\Lambda}\right)\right)\frac{\Delta_p^2}{\Lambda^2}+\ldots\right)
\end{eqnarray*}
Both equations yield the same critical value for the coupling constant. For $g>3(\hbar v)^3/2\pi\Lambda^2$ there exists charge density wave instability within mean field. Crucially the CDW instability does not gap out the Weyl nodes.

\begin{figure}
\includegraphics[width=0.5\columnwidth, clip]{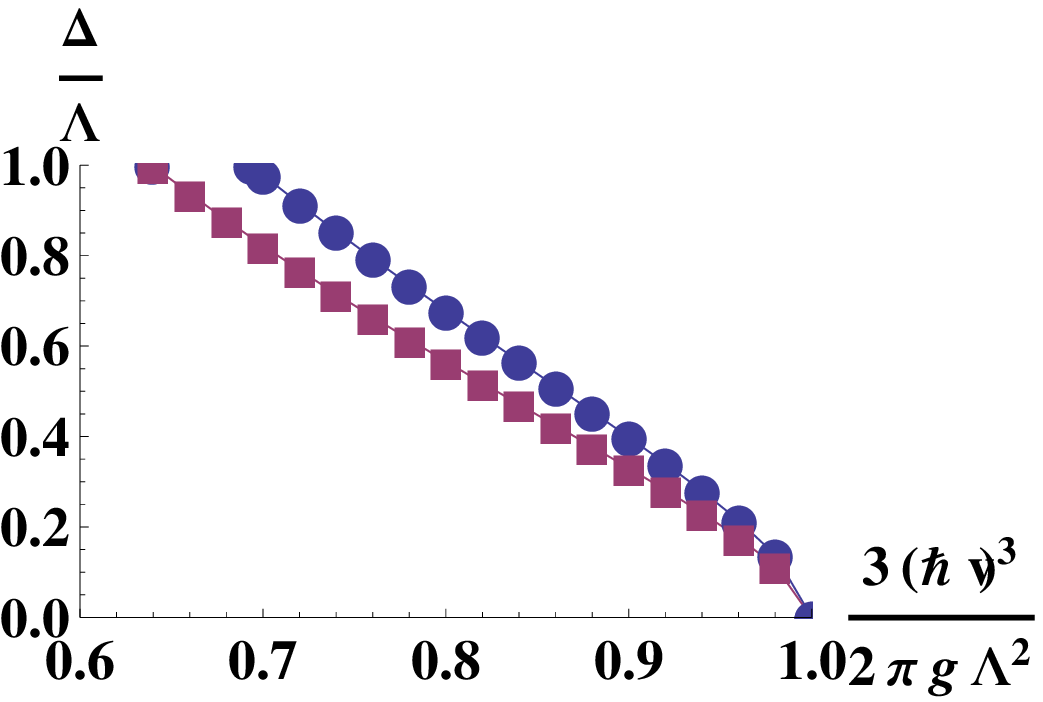}\hfill
\includegraphics[width=0.5\columnwidth, clip]{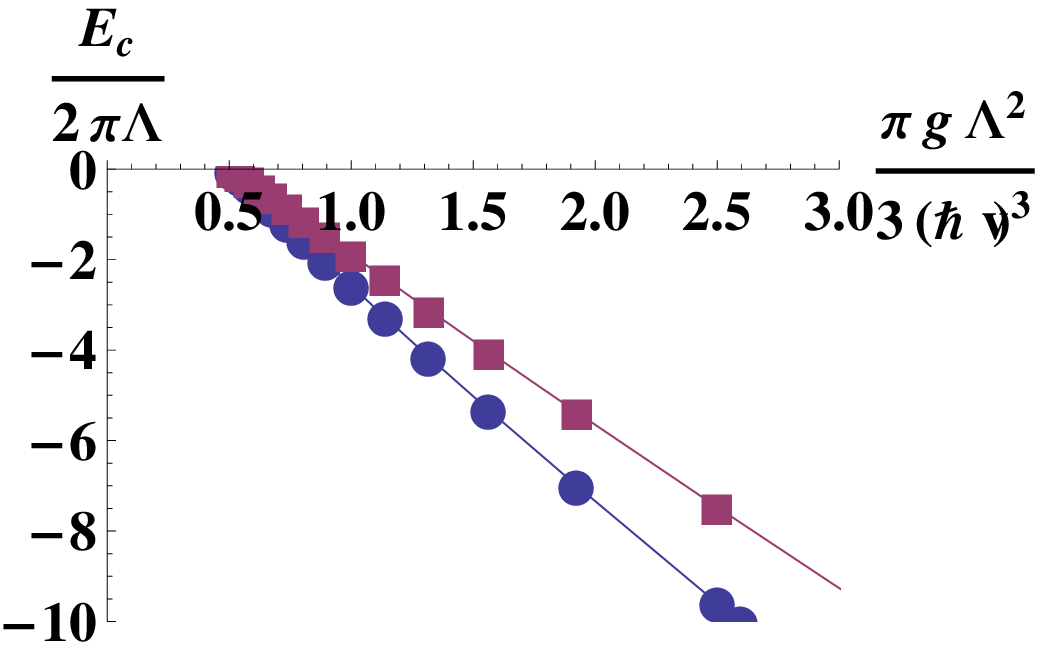}
\caption{Gap magnitude and the change in free energy of the CDW. Left: Order parameter magnitude as a function of inverse of the interaction strength for chiral (blue $\circ$) and polar (purple $\Box$) states. Right: Comparison of condensate energy $E_c$ with $\mu=0$ for aforementioned states as a function of the interaction strength.}
 \label{compenergy}
\end{figure}

In Fig.\ref{compenergy} the zero temperature free energy difference between condensate and normal states, denoted as $E_c$ or condensation energy in the figure, and magnitude of the order parameter of the two states are plotted as a function of the interaction strength. As noted analytically, the two have the same instability threshold. The magnitude of the order parameter is larger for the chiral state as compared to the polar state for the same interaction strength, leading to a greater (more negative) condensation energy for the former. Thus in this sector the lowest energy state is the chiral CDW.

The same conclusion is reached by studying the Ginzburg Landau (GL) theory, for a single order parameter, near $T_{c}$. For $T\simeq T_c$ the free energy difference is
\begin{eqnarray}
F_{CN}=-a|\Delta|^2+\frac{b}{2}|\Delta|^4+\ldots
\end{eqnarray}
For the chiral CDW phase the coefficients are
\begin{eqnarray}\label{cof}
&&a\simeq\frac{4\pi}{3k_B T}\left(\frac{1}{\beta_c\hbar v}\right)^3\left(1-\frac{T}{T_c}\right)\\\nonumber
&&b\simeq\frac{8\pi}{15(\hbar v)^3}\left(\frac{\beta_c\Lambda}{2}\ln 2+\frac{1}{4}\left(\ln\left(\frac{\beta_c\Lambda}{2\pi}\right)-\Psi\left({1\over 2}\right)\right)\right)
\end{eqnarray}
The coefficient $a$ in Eq.(\ref{cof}) is to leading order in $(\beta_c\Lambda)^{-1}$ and $\Psi(x)$ is the digamma function.  The critical temperature $T_c$, evaluated by taking $\Delta\rightarrow 0$ at $T=T_c$ in Eq.(\ref{gap}), is $k_B T_c\simeq\sqrt{\frac{3}{\pi^2}(\Lambda^2-\frac{3(\hbar v)^3}{2\pi g})}$.  For the polar CDW phase the coefficient $a$ is the same and $b$ is $3/2$ times the corresponding coefficients in Eq.(\ref{cof}). Thus the critical temperature $T_c$ is the same as the chiral CDW phase. The coefficient of the quartic term is smaller for the chiral phase and hence it has the larger order parameter and lower free energies. The most stable state in the inter nodal sector is the chiral charge density wave.

\begin{figure}
\includegraphics[width=0.5\columnwidth, clip]{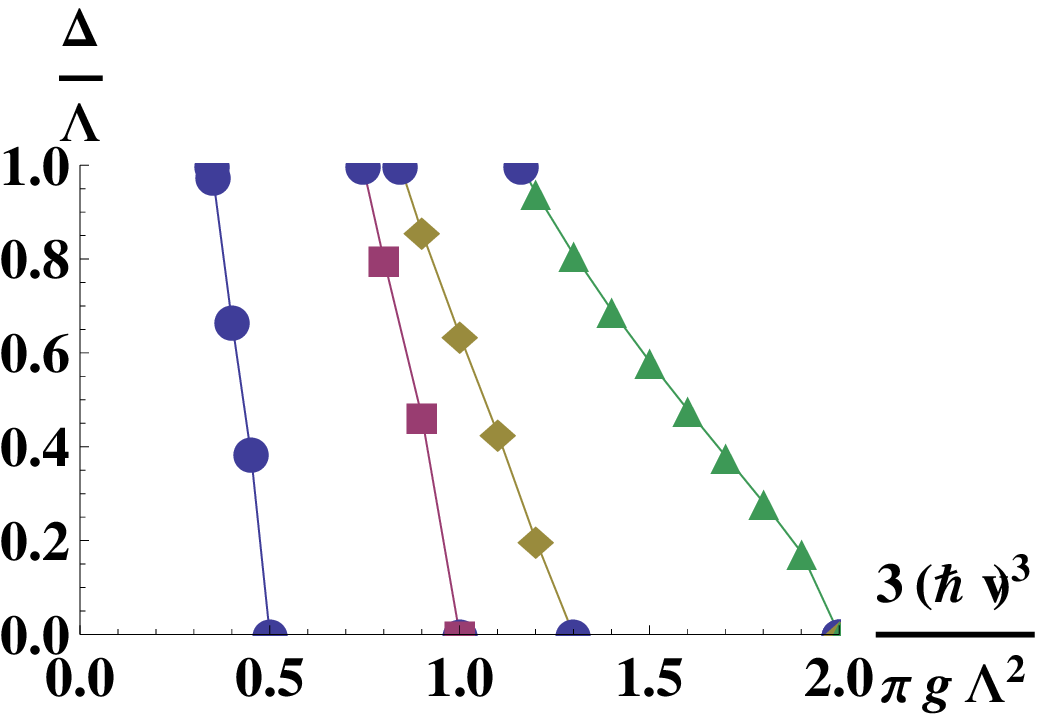}\hfill
\includegraphics[width=0.5\columnwidth, clip]{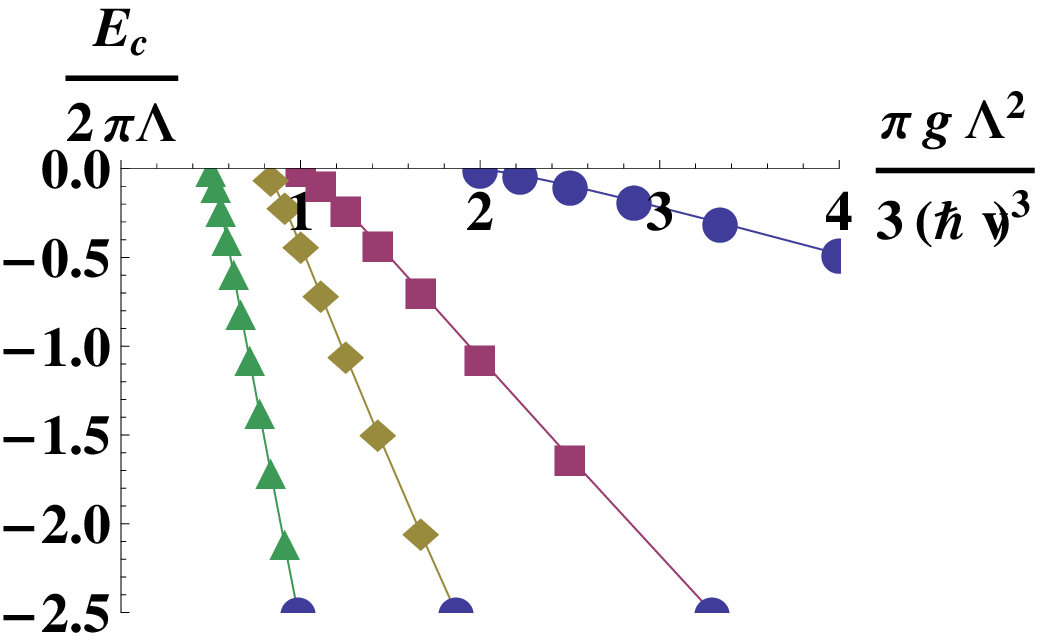}
\caption{Left: Order parameter magnitude as a function of inverse of the interaction strength for Polar-x EI $\Delta_{px1}$ (blue $\circ$), Chiral-z EI $\Delta_{cz1}$ (purple $\Box$), Chiral-x EI $\Delta_{cx1}$ (brown $\Diamond$), and Polar-z EI $\Delta_{pz1}$ (green $\triangle$). Right: Comparison of condensate energy $E_c$ with $\mu=0$ for aforementioned states as a function of the interaction strength.}
 \label{compenergy2}
\end{figure}

\noindent\underline{\textit {Intra-nodal Excitonic Insulator}} : Having established the instabilities in the inter-nodal sector we turn to those promoted by the first two terms in Eq.(\ref{int}). For $V(\vec{q}) = {g\over\Omega}$ we get
\begin{eqnarray}
V &=& -{g\over {2\Omega}}\sum_{\vec{k},\vec{k}'}\left(\vec{\Phi}_{\vec{k}}^{1\ast}\cdot\vec{\Phi}_{\vec{k}'}^{1}+\vec{\Phi}_{\vec{k}}^{2\ast}\cdot\vec{\Phi}_{\vec{k}'}^{2}\right)\\ \nonumber
\vec{\Phi}_{\vec{k}}^{1}&=& \hat{e}_{\vec{k}}^{1}\left(c_{\vec{k},+}^{L\dag}c_{k,-}^{L}+ c_{\vec{k},+}^{R\dag}c_{k,-}^{R}\right)\\ \nonumber
\vec{\Phi}_{\vec{k}}^{2}&=& \hat{e}_{\vec{k}}^{2}\left(c_{\vec{k},+}^{L\dag}c_{k,-}^{L}-c_{\vec{k},+}^{R\dag}c_{k,-}^{R}\right)
\end{eqnarray}
There are four possible particle hole instabilities for order parameters with $\hat{e}_{\vec{k}}^{1}$ component: i) Chiral-z EI: $\vec{\Delta}_{cz1}^L+\vec{\Delta}_{cz1}^R=\frac{g}{2\Omega}\langle \sum_{\vec{k}}\hat{e}_{\vec{k}}^{1}(c_{\vec{k},-n}^{L\dag}c_{k,n}^{L}+c_{\vec{k},-n}^{R\dag}c_{k,n}^{R})\rangle=(\Delta_{cz1}^L+\Delta_{cz1}^R)(\frac{\hat{x}+i\hat{y}}{\sqrt{2}})$, ii) Polar-z EI: $(\Delta_{pz1}^L+\Delta_{pz1}^R)\hat{z}$, iii) Polar-x EI: $(\Delta_{px1}^L+\Delta_{px1}^R)\hat{x}$, and iv) Chiral-x EI: $(\Delta_{cx1}^L+\Delta_{cx1}^R)(\frac{\hat{y}+i\hat{z}}{\sqrt{2}})$. This is related to the fact that $\hat{e}_{\vec{k}}^{1}$ spans only the southern hemisphere and makes the $\hat{z}$ different from the other two axes. The zero temperature order parameter magnitude, $\Delta_\alpha=2\Delta_\alpha^{R/L}$, where $\alpha$ denotes the four possible phases, is determined by the same minimization condition as in Eq.(\ref{gap}). For $g>3(\hbar v)^3/2\pi\Lambda^2$ there exists polar-z EI which has the largest gap for the same interaction strength among these four states. At zero temperature we get $\Delta_{pz1}>\Delta_{cx1}>\Delta_{cz1}>\Delta_{px1}$ and $E_c^{pz1}<E_c^{cx1}<E_c^{cz1}<E_c^{px1}$ for the same interaction strength $g$ larger than $6(\hbar v)^3/\pi\Lambda^2$ as shown in Fig.\ref{compenergy2}. By comparing it with Fig.\ref{compenergy} we see the polar-z EI and the chiral CDW are equally energetically favorable among the six possible states aforementioned for a given interaction strength. The polar-z EI state has Ginzburg Landau parameter $a$ and $b$ as $\frac{1}{2}$ and $\frac{1}{4}$ of the chiral CDW state, giving rise to smaller order parameter ($1/\sqrt{2}$ that of the chiral CDW) but the same free energy as chiral CDW at finite temperature. None of these phases gap out the Weyl node.

\begin{figure}
\includegraphics[width=0.5\columnwidth, clip]{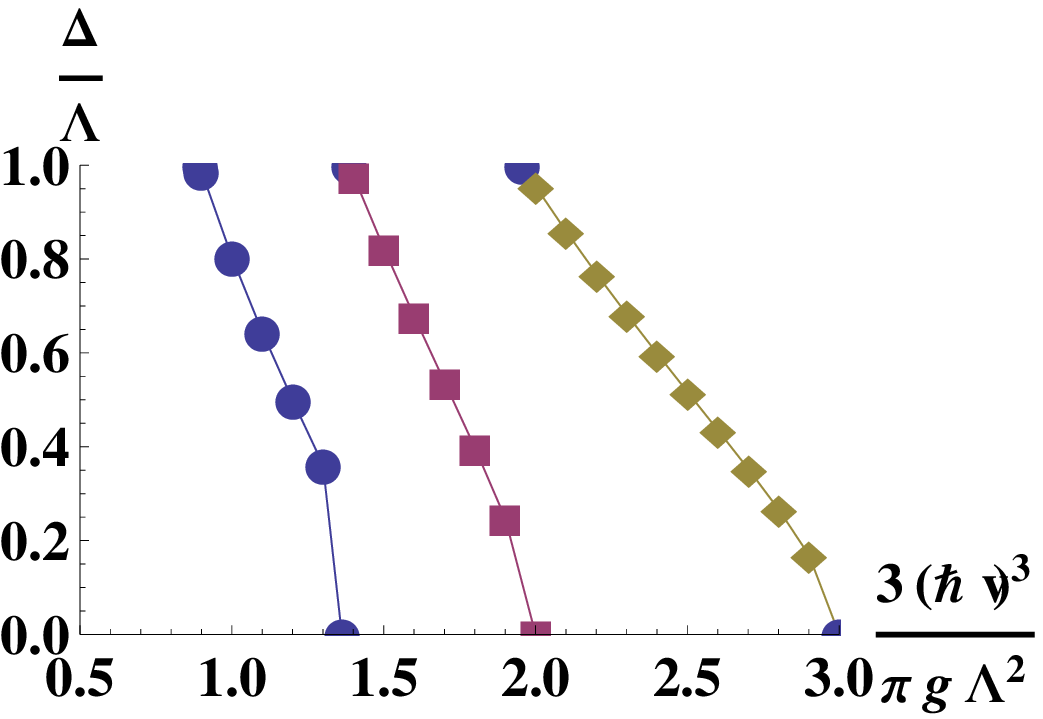}\hfill
\includegraphics[width=0.5\columnwidth, clip]{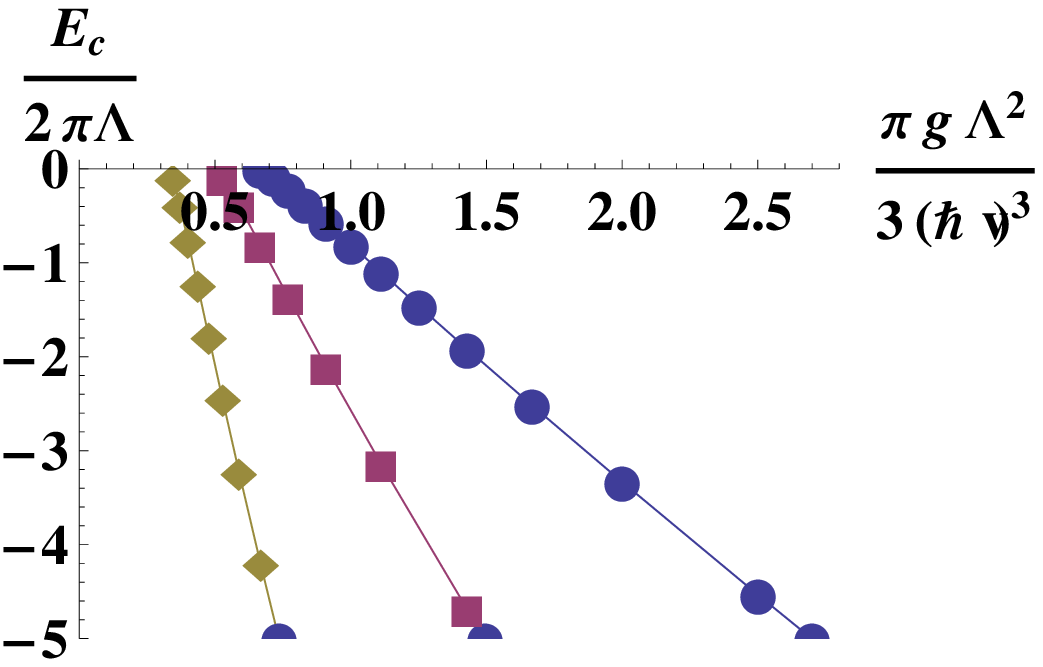}
\caption{Left: Order parameter magnitude as a function of inverse of the interaction strength for Polar EI $\Delta_{p2}$ (blue $\circ$), Chiral CDW $\Delta_{c}$ (purple $\Box$), and Chiral EI $\Delta_{c2}$ (brown $\diamond$). Chiral CDW is shown here for comparison. Right: Comparison of condensate energy $E_c$ with $\mu=0$ for aforementioned channels as a function of the interaction strength. Chiral EI is the most energetically favorable among all states.}
 \label{compenergy3}
\end{figure}

For order parameter of particle hole instabilities along $\hat{e}_{\vec{k}}^{2}$ component the order parameter have different signs between the two Dirac nodes and there are two possible EI states based on symmetry (no z-component in $\hat{e}_{\vec{k}}^2$): i) Chiral EI:
$\vec{\Delta}^L-\vec{\Delta}^R=\frac{g}{2\Omega}\langle \sum_{\vec{k}}\hat{e}_{\vec{k}}^{2}(c_{\vec{k},-n}^{L\dag}c_{k,n}^{L}-c_{\vec{k},-n}^{R\dag}c_{k,n}^{R})\rangle=(\Delta_{c2}^L+\Delta_{c2}^R)(\frac{\hat{x}+i\hat{y}}{\sqrt{2}})$ with the signs fixed by setting $\Delta_{c2}^R=\Delta_{c2}^L\equiv\Delta_{c2}/2>0$ under the assumption of inversion symmetry, and ii) Polar EI: ($\Delta_{a2}^L+\Delta_{a2}^R)(\hat{x})$ with the same sign convention. At zero temperature, the chiral EI gap equation Eq.(\ref{gap}) yields the condition
\begin{eqnarray*}
1\simeq \frac{\pi g \Lambda^2}{(\hbar v)^3}\left(1+\left(1+\ln\left(\frac{\Delta_{c2}^2}{8\Lambda^2}\right)\right)\frac{\Delta_{c2}^2}{4\Lambda^2}+\ldots\right)
\end{eqnarray*}
For $g>(\hbar v)^3/\pi\Lambda^2$ there exists a chiral EI instability with a uniform gap. From Fig.\ref{compenergy3} we see this state has the largest gap value compared with all other state for a given interaction strength. The chiral EI state is also more energetically favorable at zero temperature compared with the chiral CDW or polar-z EI. It is the only state that opens a gap at the Weyl nodes.

The chiral EI states mix particle and hole states which have opposite spin orientations in the non-interacting limit. Thus the ground state no longer has the spins aligned with their momenta. To evaluate the nature of the spin configuration we compute the expectation value of spin at momenta $\vec{k}$ for the occupied band. The result is that there exists a net polarization at each Weyl node. To understand the origin of this effect we first rotate the mean field Hamiltonian back to the $\psi_{k,\sigma}$ basis. It takes the form

\begin{equation}
H_{\pm} = H_{0\pm} - \sum_{\vec{k}}\psi_{\vec{k}\alpha}^{\dagger} \vec{\Delta}' \cdot \left[\hat{e}_{\vec{k}}^{2}\hat{e}_{\vec{k}}^{2} \mp \left(\hat{e}_{\vec{k}}^{2}\times\vec{n}\right)\hat{e}_{\vec{k}}^{1}\right]\cdot\vec{\sigma}_{\alpha\beta}\psi_{\vec{k}\beta}
\end{equation}
where $\vec{\Delta}' = \tilde{\Delta}\sin\left(\chi\right)\hat{l}+\tilde{\Delta}\cos\left(\chi\right)\hat{m}$, $\tilde{\Delta}$ is $|\vec{\Delta}_{c2}\cdot\hat{e}_{\vec{k}}^2|$, and $\chi$ is the corresponding phase. Under inversion $\hat{e}_{\vec{k}}^{1}$ does not change sign but $\hat{e}_{\vec{k}}^{2}$ does. Since we also go from one Weyl node to the other under inversion, the Hamiltonian preserves the symmetry. Averaging over polar and azimuthal angle the Hamiltonian is $\vec{\tilde{\Delta}}\cdot\vec{\sigma}/2$ at both nodes. This is the origin of the magnetization in the chiral EI state and serves as  a diagnostic of the state. An external magnetic field couples linearly to the order parameter leading to a lowering of the critical coupling, provided that it is smaller than magnitude of the gap and no Landau levels are formed.

\noindent\underline{\textit {Discussion :}}  The interplay of strong spin-orbit coupling and repulsive interaction leads to novel excitonic phases in Weyl semi-metals. For the simple case of two Weyl nodes there are both intra-nodal and inter-nodal instabilities that can be realized. For short range interaction the fully gapped chiral excitonic insulator has the lowest free energy. The p-wave nature leads to vectorial order parameters. Crucially one can define three orthonormal vectors each of which couples to three distinct order parameters namely CDW, inversion preserving EI and inversion violating EI. In analogy with superfluid liquid He, the finite temperature phase diagram will be determined by the interplay between the various sectors, which has been ignored our GL theory for single order parameters. In addition the phase of the EI order parameter determines the direction of magnetization at each node. Thus vortices have a magnetic character resulting in spin textures. The precise nature of these excitations and nature of the phase diagram, both for local interactions and long range Coulomb interactions, is currently under investigation. In particular, in analogy with $^{3}$He A, it is of great interest to determine if any of the nodal excitonic phases are possible at finite temperatures.

Our analysis focussed on systems with two Weyl nodes. The TNI heterostructures\cite{Balents1} offer a simple realization where one can look for the excitonic phases reported here. PIs are another system where, in a regime of intermediate correlations, a Weyl semi-metal with 24 nodes is conjectured to occur\cite{Savrasov}. This implies that each Weyl node can couple to 12 other nodes with opposite chirality. On the other hand the available phase space and the cut off $\Lambda$ will scale down. Whether the parameters end up being favorable to obtain the state is an open question that can only be answered once a Weyl semi-metallic state is established and characterized. Nevertheless the on going efforts to investigate systems with strong spin orbit interactions promises to focus interest in exploring possible new correlated phases of matter.

\noindent\underline{\textit {Acknowledgment :}}
The authors acknowledge the financial support by University of California at Riverside through the initial complement.

\end{document}